%Paper: hep-th/9305051
%From: Boguslaw Broda <PTBB@IBM.RZ.TU-CLAUSTHAL.DE>
%Date: Wed, 12 May 93 17:21:57 MET

%""""""""""""""""""plaintex""""""""""""""""""""""B.Broda""""
% Chern-Simons theory on an arbitrary manifold via surgery,
% by B.Broda (U.Clausthal and U.Lodz), 8 pages,
% plaintex (magstep 1),
% supported by CEC, KBN and UL
%
%""""""""definitions"""""""""macros"""""""""parameters""""""

\magnification \magstep1

\hoffset 1.5truecm
\hsize 16truecm \vsize 23.5truecm
\baselineskip 20pt \parskip5pt
\raggedbottom

\def\cM{{\cal M}}
\def\P{{\rm P}}
\def\ie{i.\thinspace e.\ }
\def\tr{{\rm Tr}}

\def\clap#1#2{
\setbox0=\hbox{$\displaystyle#2$}
\hbox to \wd0 {\hss$#1$\hss}
\kern-\wd0\hbox to\wd0{\hss$\displaystyle#2$\hss}}

\newcount\currentnumber
\def\currentno{\global\advance\currentnumber by 1
\the\currentnumber}

\def\heading#1{\bigskip\bigskip\penalty-100
\noindent{\bf\currentno.~#1}\par}

%""""""""""""""""""text starts here"""""""""""""""""""""""""

\rightline{hep-th/9305051}
\vfill
\centerline{\bf Chern-Simons theory on an arbitrary manifold
via surgery}\smallskip
\centerline{Bogus\l aw Broda}
\smallskip
{\it
\centerline{Arnold Sommerfeld Institute for Mathematical Physics}
\centerline{Technical University of Clausthal, Leibnizstra\ss e 10}
\centerline{D-W--3392 Clausthal-Zellerfeld, Federal
Republic of Germany%
\footnote{$^\dagger$}{\rm E-mail address:
ptbb@ibm.rz.tu-clausthal.de}}
\smallskip
\centerline{\rm and}
\smallskip
\centerline{Department of Theoretical Physics, University
of \L\'od\'z}
\centerline{Pomorska 149/153, PL--90-236 \L\'od\'z, Poland\/%
            \footnote{$^\ddagger$}{\rm Permanent address.}}
}
\vfill

A general formula for physical observables in Chern-Simons
theory with an arbitrary compact Lie group $G$, on an
arbitrary closed oriented three-dimensional manifold $\cM$
is derived in terms of vacuum expectation values of Wilson
loops in ${\cal S}^3$.  Surgery presentation of $\cM$ and
the Kirby moves are implemented as the main ingredients of
the approach.  The case of $G={\rm SU}(n)$ is explicitly
calculated.

\vfill
\centerline{MAY 1993}
\vfill\eject

\heading{Introduction}
Chern-Simons theory [1] is still a source of hopes and
inspiration in both mathematics [2] (low-dimensional
topology) and physics [3]. In this letter, we aim to {\it
combinatorially} derive a general formula for physical
observables (means of Wilson loops) in Chern-Simons theory
with an arbitrary compact, simple, (gauge) Lie group $G$,
on an arbitrary closed, connected, oriented
three-dimensional manifold $\cM$. The most non-trivial part
of our task consists in proper encoding topological
information about $\cM$ into the Chern-Simons path
integral; the most convenient and elegant way to do it is
seemingly the use of the {\it surgery prescription} [4].
Thus, our final formula will be expressed in terms of
vacuum expectation values of Wilson loops in the
three-dimensional sphere ${\cal S}^3$, where some of these
Wilson loops will directly correspond to ``physical
observables'', whereas some of them (``non-physical'' ones)
will serve as a presentation of topology of $\cM$ in terms
of surgery (compare with [5]).  The ``non-physical'' Wilson
loops will enter the theory through some specific linear
combinations (``surgery loops'') being a subject of a new
kind of (topological) symmetry. We will call this symmetry
the {\it Kirby symmetry}, as it corresponds to the
invariance of the surgery loops with respect to the {\it
second Kirby move} [4].  From technical point of view, one
should calculate the coefficients defining the surgery
loops in terms of ordinary (``non-physical'') Wilson loops,
taking into account the restrictions coming from the Kirby
symmetry. It appears that the non-zero coefficients we are
looking for are essentially uniquely given by the {\it
quantum dimensions} of irreducible representations of $G$.
The coefficients in the case of $G={\rm SU}(n)$, $n\geq3$,
are explicitly calculated.

The proposed purely formal ideology can be supplemented
with a more common interpretation in terms of particle
physics. First of all, one should note that the surgery
loop is nothing but an average of few Wilson loops
corresponding to virtual ``mesons'' of different
``isospins'', weighted by quantum dimensions. Thus, the
change of topology amounting to adding a two-handle is
induced by a (particle) ``virtual process''. Alternatively,
the three-dimensional sphere ${\cal S}^3$ in the presence
of the virtual process is indistinguishable from the
surgered sphere, at least as far as Chern-Simons
interactions are concerned. This way, we obtain a very
simple topology generating mechanism.

\heading{General formula}
The vacuum expectation value of the observable $\cal O$ in
Chern-Simons theory with a compact (gauge) Lie group $G$,
on ${\cal S}^3$ is formally defined by the path integral
[1]
$$
\left<{\cal O}\right>
=\int {\cal O} e^{ikS_{\rm CS}(A)}D{\cal A},
$$
where
$$
S_{\rm CS}(A)
={1\over4\pi}\int_{{\cal S}^3}\tr\left(AdA
+{2\over3}A^3\right)
$$
is the (properly normalized) Chern-Simons action functional
of the gauge field $A$, the symbol $D\cal A$ represents
Feynman's integral over all gauge orbits, and $k\in {\bf
Z}^+$ is the {\it level}. For instance, for Wilson loops in
the fundamental representation of ${\rm SU}(2)$,
``$\left<\,\right>$'' is the well-known {\it Kauffman
bracket} (incidentally, in mathematical literature, denoted
with the same symbol [6]); fundamental representations of
${\rm SU}(n)$, $n\geq3$, correspond to {\it regular}
(non-ambient) relatives of the {\it Homfly polynomial}. The
standard definition of the Wilson loop round the {\it
geometrical loop} $\cal C$ reads
$$
W_\mu = \tr_\mu {\rm P} \exp \oint_{\cal C} A,
$$
where $\mu$ numbers irreducible representations of $G$.

Now, we shall define the following ``tensors'', needed
for our further calculations:
$$
d_\mu=\left<\bigcirc_\mu\right>,
\eqno{(1)}
$$
$$
x_{\mu\nu}=\left<\cdots\clap{\bullet}{\bigodot}_\mu
\bigcup\nolimits_\nu\cdots\right>,
\eqno{(2)}
$$
$$
y_{\mu\nu\lambda}=y_{\nu\mu\lambda}
=\left<\cdots\clap{\odot}{\bigodot}_{\mu\nu}
\bigcup\nolimits_\lambda\cdots\right>,
\eqno{(3)}
$$
where ``$\circ$'' denotes a trivial (untwisted) Wilson
loop, ``$\odot$'' a trivial but non-contractible Wilson
loop in an annulus (a bordered exterior of ``$\bullet$''),
in general, non-trivially immersed in ${\cal S}^3$,
``$\clap{\circ}{\odot}$'' two parallel Wilson loops in the
annulus, ``$\bigcup$'' a piece of a Wilson loop, and
finally ``$\cdots$'' represents the rest that will be
always identical on the both sides of any equation.

A particularly useful version of the {\it satellite
formula} [7] is given by
$$
y_{\mu\nu\lambda}=\sum_\rho
c_{\rho\mu\nu} x_{\rho\lambda},
\eqno{(4)}
$$
where $c_{\rho\mu\nu}=c_{\rho\nu\mu}$ are the
{\it Clebsch-Gordan coefficients} corresponding to the
decomposition of the representations inside the annulus
$$
R_\mu\otimes R_\nu=\bigoplus_\rho c_{\rho\mu\nu} R_\rho.
$$

In our notation, the Kirby symmetry assumes the
following simple graphical form
$$
\left<\cdots\clap{\bullet}{\bigodot}_\omega
\bigcup\nolimits_\Omega\cdots\right>
=\left<\cdots\clap{\odot}{\bigcup^\omega}_\Omega
\cdots\right>,
\eqno{(5)}
$$
where the loops denote now surgery loops rather than the
Wilson ones, \ie
$$
W_\omega(A)=\sum_\mu \Delta_\mu W_\mu(A),
\qquad
W_\Omega(A)=\sum_\mu \delta_\mu W_\mu(A),
\eqno{\rm (6a,b)}
$$
and $\Delta_\mu$, $\delta_\mu$ are the coefficients we are
looking for. Applying the formula for a {\it connected
sum} [7] to (3) yields
$$
y_{\mu\nu\nu}
=\left<\cdots\clap{\odot}{\bigodot}_{\mu\nu}
\bigcup\nolimits_\nu\cdots\right>
=d_\nu\left<\cdots\clap{\odot}{\bigcup^\mu}_\nu\cdots\right>.
\eqno{(7)}
$$
By virtue of (6), (7) and (4), we can rewrite RHS of (5) in
the following manner
$$
\left<\cdots\clap{\odot}{\bigcup^\omega}_\Omega\cdots\right>
=\sum_{\mu,\nu} \Delta_\mu\delta_\nu d_\nu^{-1}
\sum_\rho c_{\rho\mu\nu} x_{\rho\nu},
\eqno{(8)}
$$
whereas LHS of (5) is of the form
$$
\left<\cdots\bigodot\nolimits_\omega
\bigcup\nolimits_\Omega\cdots\right>
=\sum_{\mu,\nu} \Delta_\mu \delta_\nu x_{\mu\nu}.
\eqno{(9)}
$$
Since eq.~(5) should be satisfied for any observables (and
hence for any $x_{\mu\nu}$), we have
$$
\Delta_\mu\delta_\nu
=\sum_\rho \Delta_\rho \delta_\nu d_\nu^{-1}
c_{\mu\rho\nu}.
\eqno{(10)}
$$
Finally (compare with ref.~[8]),
$$
d_\nu \Delta_\mu
=\sum_\rho c_{\mu\rho\nu} \Delta_\rho,
\eqno{(11)}
$$
where we have tacitly assumed that $d_\nu \not=
0$ (in (8)), as well as $\delta_\nu \not= 0$ (in (10)).

We should note that $d_\nu$ can be identified
to the quantum dimension of the representation $\nu$
[9], and the ``classical'' {\it deformation parameter}
expressed by the $k^{\rm th}$ primitive root of unity,
where $k$ is the level, undergoes a ``quantum'' shift [1]
corresponding to
$$
k\longrightarrow r=k+h,
\eqno{\rm(12a)}
$$
where $h$ is the {\it dual Coxeter number}, \ie
$$
q_{\rm cl}=e^{2\pi i\over k}
\longrightarrow q=e^{2\pi i\over r}.
\eqno{\rm(12b)}
$$

The solution of eq.~(11) in the form $\Delta_\nu = d_\nu$
immediately follows from the locality of the theory, and
from the satellite formula applied to a trivial Wilson
loop, \ie
$$
\left<\bigcirc_\nu\right> \left<\bigcirc_\mu\right>
=\left<\clap{\circ}{\bigcirc}_{\nu\mu}\right>
=\sum_\rho c_{\rho\nu\mu}\left<\bigcirc_\rho\right>,
\eqno{\rm(13a)}
$$
or by virtue of (1)
$$
d_\nu d_\mu = \sum_\rho c_{\rho\nu\mu} d_\rho,
\eqno{\rm(13b)}
$$
as well as from the character formula for the quantum
dimension and its reality ($d_\mu^* = d_\mu$) for $|q|=1$
[10].

Since the roles of $\omega$ and $\Omega$ in (5) could be
interchanged ($\omega\leftrightarrow\Omega$) from the point
of view of the Kirby symmetry, and we can build an
arbitrary representation of the {\it primitive} ones, we
should put $\nu \in \P$ (the set of primitive representations
of $G$) in (11) to assure the consistency. Then
$$
\delta_\nu=\cases{1,&if $\nu \in \P$,\cr
0,&otherwise,\cr}
$$
and $d_\nu \not= 0$ ($\nu\in\P$).

The observation that $\nu$ should be restricted to
primitive representations of $G$ is crucial for two
reasons: (i) eq.~(11) becomes ``irreducible'', \ie there
are no (and there should not be [11]) subsets of non-zero
$\Delta$'s not connected by (11), and hence the solution
$\Delta_\mu = d_\mu$ is essentially unique; (ii) eq.~(11),
looked at as a recursive relation, terminates yielding a
finite set of representations.

Thus, physical observables in Chern-Simons theory with
the (gauge) Lie group $G$, on the manifold $\cal M$ are
expressed by
$$
\left<\prod_i W_{\mu_i}(A) \right>_{\cal M}
={
\left<\prod_i W_{\mu_i}(A) \prod_j W_{\omega_j}(A)\right>
\over
\left<\prod_j W_{\omega_j}(A)\right>
},
\eqno{(14)}
$$
where $W_{\mu_i}(A)$ is a ``physical'' Wilson loop, and
$W_{\omega_j}(A)$ is
$$
W_\omega(A) = \sum_\mu d_\mu W_\mu(A),
\eqno{(15)}
$$
put on the $j^{\rm th}$ surgery loop entering the surgery
presentation of $\cal M$.

\heading{${\rm SU}(n)$ example}
According to [9], the quantum dimension of the
representation corresponding to the highest weight $\mu$ is
$$
d_\mu = \prod_{\alpha \in \Phi^+}
{
q^{(\mu+\rho,\alpha)/2}-q^{-(\mu+\rho,\alpha)/2}
\over
q^{(\rho,\alpha)/2}-q^{-(\rho,\alpha)/2}
},
\eqno{(16)}
$$
where $q$ is the deformation parameter (12b), $\Phi^+$ is
the set of the positive roots, and $\rho$ is the {\it Weyl
vector} (half the sum of the positive roots).

As an explicit example, we will consider the case of $G =
{\rm SU}(n)$, $n \geq 3$. From eq.~(16) it follows that
$$
d_{\mu_1 \mu_2 \dots \mu_{n-1}}
= \prod_{1 \leq i < j \leq n}
{
q^{\left(\sum_{l = i}^{j-1} \mu_l +j-i \right)/2}
-q^{-\left(\sum_{l = i}^{j-1} \mu_l +j-i \right)/2}
\over
q^{j-i \over 2} - q^{i-j \over 2}
},
\eqno{(17)}
$$
with $\mu_l$ the Dynkin indices corresponding to $\mu$. It
is obvious that $d_{\mu_1 \mu_2 \dots \mu_{n-1}}$ vanishes
the first time, truncating (11), for
$$
\sum_{l = 1}^{n-1} \mu_l
= r+1-n = k+h+1-n = k+1,
\eqno{(18)}
$$
where eq.~(12) as well as the fact that the dual Coxeter
number $h=n$ for $G={\rm SU}(n)$ have been taken into
account.  Thus, only the representations corresponding to
the Young diagrams contained in the rectangle with the
sides of the length $k$ and $n-1$ enter (15). Then, the
recursive relation (11) terminates, and consequently
$$
W_\omega(A)
=\sum_{\scriptstyle \mu_1,\mu_2,\dots,\mu_{n-1} \geq 0
\atop \scriptstyle \sum_{l = 1}^{n-1} \mu_l \leq k}
d_{\mu_1\mu_2\cdots\mu_{n-1}}
W_{\mu_1\mu_2\cdots\mu_{n-1}}.
\eqno{(19)}
$$

The expression (14) for $G={\rm SU}(2)$, modulo the
normalization (corresponding to the first Kirby move), is
known in literature as the invariant of Reshetikhin, Turaev
and Witten.

\bigskip\bigskip\noindent
{\bf Acknowledgements}

The author is indebted to Prof.~H.-~D.~Doebner for his kind
hospitality in Clausthal, and to Dr.~B.~Jur\v co for
sending ref.~[10]. The work was supported by the CEC grant
CIPA3510PL921596, the KBN grant 202189101, and the
University of
\L\'od\'z grant.

\vfill\eject

\frenchspacing

\item{[1]} E. Witten, Commun. Math. Phys. 121 (1989) 351.

\item{[2]} M. Atiyah, {\it The geometry and physics of
knots} (Cambridge University Press, Cambridge, 1990).

\item{[3]} D. Birmingham, M. Blau, M. Rakowski and G.
Thompson, Phys. Reps. 209 (1991) 129.

\item{[4]} L. H. Kauffman, {\it Knots and Physics} (World
Scientific, Singapore, 1991), Part~I, Chapt.~16, and
references therein.

\item{[5]} E. Guadagnini, Nucl. Phys. B 375 (1992) 381.

\item{[6]} B. Broda, J. Phys. A, to be published (1993).

\item{[7]} E. Guadagnini, Int. J. Mod. Phys. A 7 (1992)
877.

\item{[8]} K. Walker, {\it On Witten's 3-manifold Invariants},
preprint (1990).

\item{[9]} R. B. Zhang, M. D. Gould and A. J. Bracken,
Commun. Math. Phys. 137 (1991) 13.

\item{[10]} J. Fuchs, {\it Affine Lie Algebras and Quantum
Groups} (Cambridge University Press, Cambridge, 1991),
Chapt.~4.

\item{[11]} B. Broda, {\it Chern-Simons approach to
three-manifold invariants}, e-preprint hep-th/9301091, to
be published.
\bye